\documentclass[prl,twocolumn,aps]{revtex4}
\usepackage{graphicx}
\usepackage{natbib}
\usepackage{color}
\usepackage{units}
\usepackage{amsmath, amsthm, amssymb}
\usepackage[normalem]{ulem}

\def\strutdepth{\dp\strutbox}
\def\nw#1{\strut\vadjust{\kern-\strutdepth\vtop to0pt{\vss\hbox to\hsize
{\hskip\hsize\hskip5pt$\leftarrow$\hss\strut}}}{\em #1}}
\begin{document}

\title{Cusp-shaped Elastic Creases and Furrows}

\author{S. Karpitschka$^{1}$, J. Eggers$^{2}$, A. Pandey$^{1}$,
and J. H. Snoeijer$^{1,3}$}
\affiliation{
$^1$Physics of Fluids Group, Faculty of Science and Technology,
Mesa+ Institute, University of Twente,
7500 AE Enschede, The Netherlands.\\
$^2$School of Mathematics, 
University of Bristol, University Walk,
Bristol BS8 1TW, United Kingdom  \\
$^3$Mesoscopic Transport Phenomena, Eindhoven University of Technology,
Den Dolech 2, 5612 AZ Eindhoven, The Netherlands}

\begin{abstract}
The surfaces of growing biological tissues, swelling gels, and compressed
rubbers do not remain smooth, but frequently exhibit highly localized
inward folds. We reveal the morphology of this surface folding in a novel experimental setup, 
which permits to deform the surface of a soft gel in a controlled fashion. 
The interface first forms a sharp furrow, whose tip size decreases rapidly with deformation. 
Above a critical deformation, the furrow bifurcates to an inward folded crease of vanishing tip size. 
We show experimentally and numerically that both creases and furrows exhibit a universal cusp-shape, 
whose width scales like $y^{3/2}$ at a distance $y$ from the tip. 
We provide a similarity theory that captures the singular profiles before and after the
self-folding bifurcation, and derive the length of the fold from large deformation elasticity. 
\end{abstract}

\date{\today}%

\maketitle
Compressing a slice of soft white bread, one observes the formation of
a crease, a localized indentation where part of the surface folds into
a self contact. Similar patterns appear, for instance, on the surfaces
of swelling gels~\cite{T86,TKH08}. In biology, such elastic structures
are called sulci, which  are prime morphological features
of human brains and growing tumors~\cite{HM11,DC11,HM12}.
As a result, creases have attracted considerable
attention, experimentally, theoretically, and from a numerical point of 
view \cite{DC11,Holmes10,Paulsen16,HZS09,TKH08,BC10,DW11,CCSH12a,CH12,
GL07,DB12,CCSH12b,HM12,HM11,WC14,JCHS14, Stewart16}.
Yet in spite of their ubiquity and importance, a
quantitative theoretical description of the morphology of localized
indentations is still missing. 

\begin{figure}
\centering\includegraphics[scale=0.9]{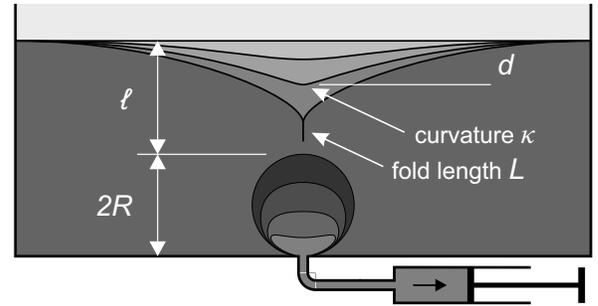}
\caption{\label{fig:setup}
The interface of a soft gel is deformed by slowly reducing the volume
of a liquid inclusion (initial radius $R$), located at an initial
distance $\ell$ below the free surface. First, a localized furrow with
tip curvature $\kappa$ forms; at larger deformations the furrow
bifurcates into a crease which folds the free surface onto itself
over a length $L$. 
}
\end{figure}

Past approaches have focused on the idealized problem
of a half-space of elastic material, which is compressed uniformly
parallel to the interface. Above a critical compression, the uniform
state becomes unstable toward sinusoidal deformation of the interface
\cite{B63}. However, since this setup lacks a characteristic length scale,
perturbations grow without bound even in the nonlinear regime. Additional
regularizing features have to be invoked, such as adding a thin film of stiff
material on the surface \cite{HM11,HM12}. 
While the metastability of smooth and creased configurations has been studied in some detail~\cite{HZS09,HM11,DB12}, 
much less is known on the profiles of localized indentations. Here we propose a new experimental
setup which guarantees the formation of a single indentation of finite size,
which bifurcates between two different structures, see Fig.~\ref{fig:setup}. 
It allows to reveal for the first time the self-similar shape properties of both structures, and
provide quantitative analytical descriptions thereof.

A highly deformable PDMS gel (Dow Corning CY52-276, components A \& B mixed 1:1, shear modulus $\mu=1$~kPa)
is prepared in a container of footprint 3x3~cm. A 10 times stiffer gel was prepared for some experiments by adding $5\%$ of Dow Corning Sylgard 184 (polymer \& curing agent mixed 10:1, yielding $\mu=11$~kPa). The gels were cured overnight at room temperature, protecting the air-exposed free surface from dust. By depositing a water drop inside the gel prior to
curing, we create a liquid inclusion of initial radius $R$ at the bottom of the container. $\ell$ denotes the initial distance between inclusion and free surface (cf. Fig.~1); we used  $\ell=1-8$~mm and $R/\ell = 0.4-6$.
Subsequently, the water is extracted slowly ($\gtrsim 100$~s for the droplet volume) through a small hole at the bottom
(cf. Fig.~\ref{fig:setup}), creating a quasi-stationary axisymmetric strain field
in the gel, and an increasingly sharp indentation of the free surface. 
A two-dimensional version of the experiment was realized by creating a
cylindrical inclusion at the bottom of the container. In that case, the
inclusion was templated by a cylinder of polyethylene glycol
($M=1000$~g/mol, melting point $37-40^\circ$) which was removed by melting
after the gel was cured.

The free surface profile was measured as a shadowgraph through a
long-distance microscope with a spatial resolution on the order of
10 microns. The deformation field and the length of the self-contacting
surface fold were determined by tracking fluorescent particles embedded
inside the gel (see supplemental material for details~\cite{sup}). 
The amplitude of the deformation is measured by $d$, the deflection of
the free surface relative to its reference level. The sharpness of
the deflection is quantified by its curvature $\kappa$ in the image plane. 

\begin{figure}
\centering\includegraphics[scale=1]{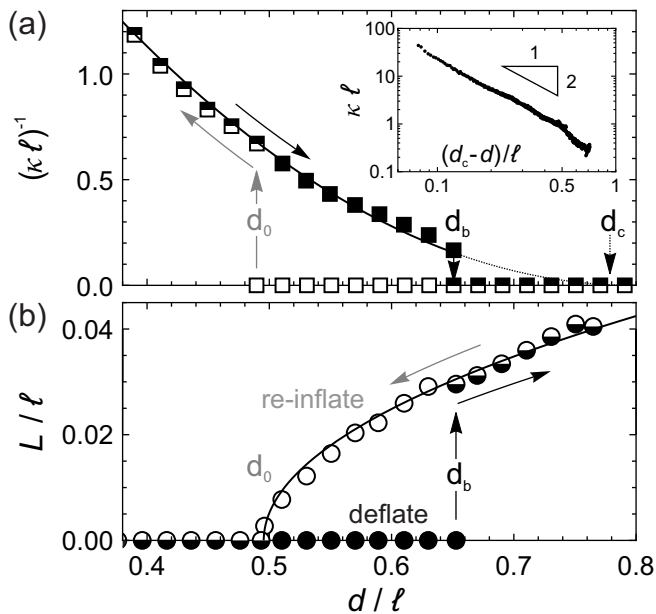}
\caption{\label{fig:states}
Bifurcation between ``furrow" and ``crease" (axisymmetric cavity, $\mu=1$~kPa, $R=2.1$~mm, $\ell=2.4$~mm). (a) Tip radius of curvature $\kappa^{-1}$ in the image plane and (b) fold length
$L$, as a function of the deformation amplitude $d$. Arrows indicate the
course of the experiment. Increasing the deformation (filled symbols) beyond $d_b$ nucleates a fold of finite length. Decreasing the deformation again (open symbols), the fold disappears continuously at $d_0$. The solid lines are~(\ref{eq:divergence}) and~(\ref{L}) for curvature and fold length, respectively. The inset demonstrates~(\ref{eq:divergence}) in a double logarithmic plot (here: $R=1.9$~mm, $\ell=2.4$~mm).}
\end{figure}

Figure~\ref{fig:states} shows the result of a typical experimental
run, obtained by first deflating the water drop to vanishing size (filled symbols), and subsequently re-inflating it up again to its original size (open symbols).
The deformation is quantified by $d/\ell$ shown on the horizontal axis.
Upon increasing deformation, the
gel develops an increasingly sharp furrow, as measured by the dimensionless
radius of curvature $(\kappa\ell)^{-1}$ (panel (a); the furrow's self-similar
shape is investigated below).
At a deformation $d = d_b$, the
furrow bifurcates toward a 
crease (similar to previously reported behavior~\cite{HM11,DB12}): part of the surface folds into a self-contact of length $L$, connected to a free-surface cusp of \emph{vanishing} tip curvature.
In the axisymmetric version of the experiment, this is
accompanied by a breaking of axisymmetry, the crease being essentially
two-dimensional.
Experimental data are fitted with the scaling law
(panel (a), solid line)
\begin{equation}
\label{eq:divergence}
\kappa = k \ell (d_c - d)^{-2},
\end{equation}
which suggests the existence of a critical deformation $d_c$ at which the
tip radius of curvature vanishes; however, this critical scaling is cut off by a
discontinuous (first order) transition toward the crease at $d = d_b$. 

Directly after formation of the crease, a self-contact of length
$L \approx 0.03\ell$ forms, while the radius of curvature of the
new structure jumps to zero. With increasing deformation, $L$ increases
further (circles).
Re-inflating the liquid inclusion again, so as to decrease $d$, $L$ decreases beyond
its original value to go to zero in a continuous fashion at another
critical value $d_0$; this is described by the critical
behavior (panel~(b), solid line),

\begin{equation}
\label{L}
L = c\sqrt{\ell(d - d_0)},
\end{equation}
reminiscent of a second order transition, to be discussed
below. Below $d = d_0$, the crease disappears and the interface shape
returns to a furrow, in the course of which the tip radius jumps to a
finite value. Decreasing the deformation further, the tip radius returns
to its original value along the same curve, indicating that the entire
process is reversible.
To check reversibility, we repeated the whole cycle several times for each specimen, which yielded nearly identical results (typical deviation $\lesssim2\%$). Merely $d_b$ was slightly smaller by about $4\%$ as compared to the first creasing event. This could indicate the formation of a localized nucleation seed due to the initial creasing~\cite{CCSH12b}.

The scaling laws Eqs.~\ref{eq:divergence} and~\ref{L} are universal features of the creasing instability. These were consistently observed, where we in total considered about 25 different configurations (axisymmetric and 2D, soft and stiff gels) with $\ell=1-8$~mm and $R/\ell = 0.4-6$. $R\ll\ell$ precludes large deformations because of the limited droplet volume. For $R\lesssim0.8\ell$, creasing was not induced before the droplet was drained completely (axisymmetric samples; $R\lesssim0.5\ell$ for 2D). Otherwise, all experiments show similar curves as in Fig.~\ref{fig:states} with universal scaling laws. As expected, the precise values of $d_b$ and $d_0$ are not universal (see supplement~\cite{sup}).

\begin{figure}
\centering\includegraphics[scale=1]{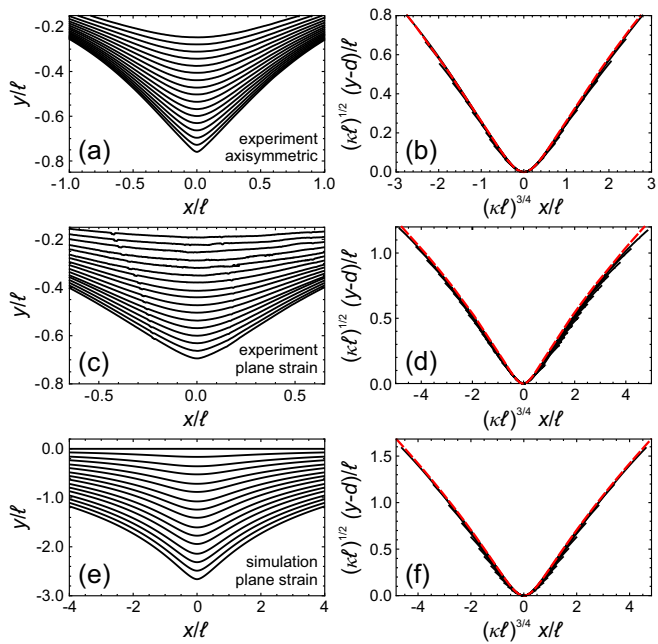}
\caption{Self-similar evolution of free surface profiles for 3D and 
2D experiments, and 2D simulations, prior to the creasing instability.
Left: measured (simulated) profiles
Right: profiles rescaled according to~(\ref{eq:collapse}), and superimposed
with the similarity solution $\Phi$ (red).
\label{fig:profiles}
}
\end{figure}

The shape of the furrow is perfectly self-similar, as is shown in
Fig.~\ref{fig:profiles}. 
To describe this self-similar structure analytically,
we hypothesize that 
the surface shape is described by a plane curve with a smooth parametric 
representation $x(s),y(s)$, which has been used successfully to describe
cusp formation quantitatively in free surface flows and in optics
\cite{EF_book,ES17}; a rigorous rationale for such a description is
provided for example by a complex mapping between the free surface and
the unit circle \cite{JM92}, as has been verified experimentally for viscous flows~\cite{LORENCEAU}. To picture cusp formation of a parameterized curve
$x(s),y(s)$ geometrically, one can imagine these components being deformed 
smoothly such that the curve self-intersects. At the point of
self-intersection, the curve is a cusp with a singular tip. Just before
intersection the curve opens into a universal smooth curve.

Namely, a critical point of the curve corresponds to $x'(0)=y'(0)=0$,
so expanding about $s=0$ to lowest non-trivial order yields \cite{EF13} 
\begin{equation}
x =\epsilon s + s^3/(2^{3/2}a), \quad y = s^2/2,
\label{eq:map}
\end{equation}
where $\epsilon=0$ corresponds to the critical (cusp) point, and $a$ is
a parameter controlling the opening of the cusp
$y = \left(ax\right)^{2/3}$. In the $x$-component
we expanded to third order, since any quadratic term can be eliminated
using $y$, implying a rotation. The curvature of~(\ref{eq:map}) at the
origin is $\kappa = \epsilon^{-2}$, so~(\ref{eq:map}) can be written in 
similarity form 
\begin{equation}
\label{eq:collapse}
y\kappa^{1/2} = \Phi\left(\xi\right), \quad \xi = x \kappa^{3/4},
\end{equation}
where $\Phi$ is defined implicitly:
$\xi^2=2\Phi\left(1+\Phi/(\sqrt{2}a)\right)^2$, see~\cite{EF13}.
As seen in Fig.~\ref{fig:profiles}~(b,d,f), the similarity
form~(\ref{eq:collapse})
is in excellent agreement with both 2D and 3D experiments and simulation, 
and the collapsed data agrees very well with the universal similarity function 
$\Phi(\xi)$. The single adjustable parameter $a$ is determined by the
outer geometry of the problem. {In addition, the relation between the
vertical deformation scale and $\kappa$ implied by~(\ref{eq:collapse})
is consistent with~(\ref{eq:divergence}). Of course, this geometric
analysis cannot describe the precise value of the tip curvature
$\kappa(d)$, which must be derived from large deformation elasticity theory
\cite{Suo_notes_fin,Suo_notes_rubber}.

To show that the observations above are well described by the mechanics of elasticity, we performed
two-dimensional (plane strain) finite element simulations implemented
in oomph-lib~\cite{oomphlib}, using the theory
of finite deformations \cite{Suo_notes_fin}, with an incompressible
neo-Hookean constitutive equation \cite{Suo_notes_rubber}.
In large deformation theory the coordinates ${\bf X}$ of the
undeformed state of the system (the reference state, see Fig.~\ref{fig:crease})
is mapped upon the 
current, deformed state of the system as ${\bf x} = {\bf f}({\bf X})$.
For a neo-Hookean elastic material, the Cauchy stress is \cite{Suo_notes_fin} 
\begin{equation}
\sigma_{ij} = \mu \frac{\partial x_i}{\partial X_k}
\frac{\partial x_j}{\partial X_k} - p\delta_{ij},
\label{nH}
\end{equation}
where $\mu$ is the shear modulus and $p$ the solid pressure (here defined up to a constant), which ensures
incompressibility: $\det(\partial x_i/\partial X_j) = 1$.
Elastic equilibrium is determined by
$\partial \sigma_{ik}/\partial x_k = 0$. For simulation details see supplement~\cite{sup}.
The result of the simulations is given in panels (e,f) of Fig.~\ref{fig:profiles}. They recover the same features as in experiments, including the similarity collapse with
the same universal shape superimposed.

\begin{figure}
\centering\includegraphics{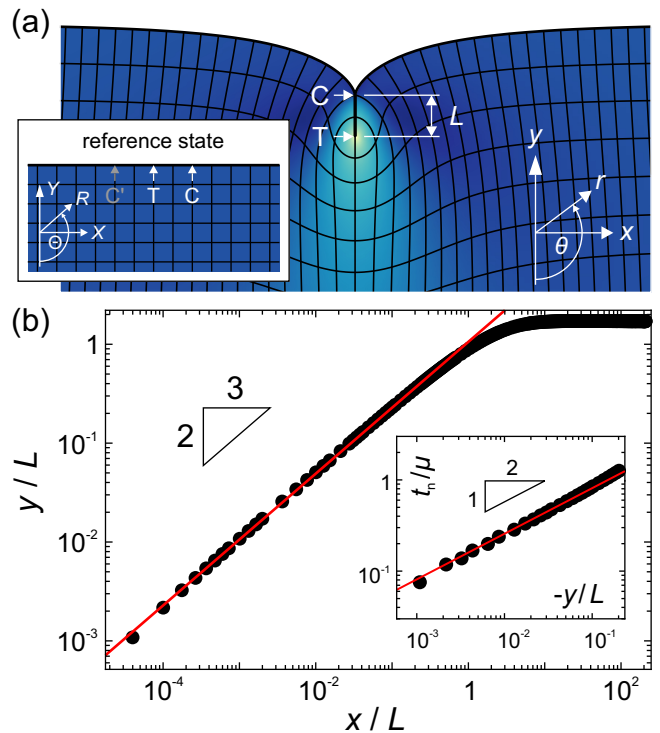}
\caption{Numerical simulation of the creased state, with self-contacting
fold of length $L$; the tip of the fold is at T, the self-contact ends
at $C$ (the tip of the cusp). 
(a) The deformed state ${\bf x}$, showing the solid pressure as a color
plot; the reference state ${\bf X}$ is shown in the inset. 
(b) Surface profile of the crease above the fold 
($x$ and $y$ measured relative to C; red line: fit with a $2/3$-power law).
The inset shows $1/2$-power law
of the normal (contact) traction near C. 
\label{fig:crease}}
\end{figure}

The remaining challenge is to understand the morphology of the creased
state, which contains two singular points, respectively indicated as
C and T in Fig.~\ref{fig:crease}~(a).
To derive a solution of the creased state, we start from the fold solution
around point T \cite{SP65,HZS09}, which maps an elastic half space onto a fold
of infinite length (coordinates are defined in Fig.~\ref{fig:crease}~(a);
here the origin lies in point T):
\begin{equation}
\theta = 2\Theta, \quad r = R/\sqrt{2}, \quad
p = -3\mu\ln r / 2,
\label{SP}
\end{equation}
which is an exact solution of~(\ref{nH}). On the 
fold ($x=0$), the principal stretches are
$\lambda_{x,y} = \sqrt{2},1/\sqrt{2}$, and the elastic free energy density is
$W = 5\mu/4$. The logarithmic divergence of $p$ near T is uncritical for a macroscopic description of the experiment since $p\lesssim 100$~kPa down to molecular length scales. To numerically simulate the creased state, we use a
large domain ($\sim 320 L$) under horizontal compression, impose
(\ref{SP}) near T, and require a non-negative normal traction and
vanishing tangential traction on TC (see supplement for details~\cite{sup}). 

Figure~\ref{fig:crease}~(a) shows the deformed computational domain, while 
(b) reveals a power-law behavior $y \sim (ax)^{2/3}$ of the interface above the
self-contact, with 4 decades of spatial resolution. Hence, the interface
forms an ideal cusp, $y \sim x^{2/3}$, which is the limiting case
$\epsilon=0$ in (\ref{eq:map}). In the supplement~\cite{sup} we also provide 
experimental evidence for this scaling. To access the surface profile near the self-contact experimentally with sufficient spatial resolution, we bent an elastic rod until it creased on its surface and recorded its shadow graph. Despite the significantly different outer geometry, we find the same 2/3 exponent for the morphology of the crease, highlighting the universality of this result.

This scaling can be derived
analytically by noting that near C, where the fold opens, the shape is
slender: $x\ll y$ (from now on we use C as the coordinate origin).
Hence deformations relative to that of the fold are small and we can
expand to linear order in the deformations $u,v$:
\begin{equation}
x = \lambda X + u(\lambda X,Y/\lambda), 
\quad y = \lambda^{-1} Y + v(X\lambda,Y/\lambda),
\label{compression_map_lin}
\end{equation}
where $\lambda$ is the stretch near C. 
We use coordinate systems as shown in Fig.~\ref{fig:crease}~(a),
but for simplicity use a reference state which is rotated clockwise 
by 90 degrees. 
As shown in \cite{B63}, if we introduce a stream function
$u = \partial_y \psi$, $v = -\partial_x \psi$, the linearized elasticity
problem reduces to
\begin{equation}
\triangle \bar{\triangle}\psi \equiv \triangle \Psi = 0,
\label{bi}
\end{equation}
where $\triangle$ and $\bar{\triangle}$ denote the Laplacian in
the deformed $(x,y)$ and reference $(X,Y)$ coordinates, respectively. 

Similar to the analysis of the cusp in a viscous fluid~\cite{EF_book},
we make the self-similar ansatz
$\psi = r^{\alpha} f(\theta)=R^{\alpha}\bar{f}(\Theta)$;
$\theta = \Theta = \pi$
corresponds to the cusp line, along which we impose vanishing shear and,
outside the self-contact, vanishing normal stress. Using that $f$ is odd,
we find from the second equation (\ref{bi}) that \cite{EF_book}
\begin{equation}
\bar{\triangle}\psi=\Psi = A r^{\alpha-2}\sin(\alpha-2)\theta
\label{Lap}
\end{equation}
and
$p = A\mu r^{\alpha-2}\cos(\alpha-2)\theta$, where $A$ is an arbitrary
constant. Now solving (\ref{Lap}) 
in reference coordinates, homogeneous solutions are
$\bar{f}_{\{1,2\}} = \{\sin,\cos\}(\alpha\Theta)$. In polar
coordinates, the transformation between deformed and reference
coordinates reads $R/r = \left(\lambda^2\cos^2(\theta) +
\lambda^{-2}\sin^2(\theta)\right)^{1/2}\equiv g$ and
$\lambda^2 \tan\Theta = \tan\theta$, and so 
\begin{equation}
f_{\{1,2\}} = \{\sin,\cos\}(\alpha\Theta) g^{\alpha}. 
\label{hom_sol}
\end{equation}

An odd particular solution of (\ref{Lap}) is found from the
standard formula as 
$f_p = A(f_1 I_2(\theta) - f_2 I_1(\theta))$, where
\begin{equation}
I_{\{1,2\}}(\theta) = 
\int_0^{\theta} \frac{\{\sin,\cos\}(\alpha\Theta)
\sin(\alpha-2)\theta}{\alpha g^{\alpha}} d\theta.
\label{integral}
\end{equation}
A general solution to (\ref{bi}) 
can be written $f = f_p + B f_1$, where the constants
$A,B$ must be chosen to satisfy $\sigma_{ij}n_j=0$ at $\theta =\pi$,
where $n_j$ is the true normal. 

For there to be a non-trivial solution, the determinant of this
system of equations must vanish, which after using that
$2\alpha(\alpha-1)I_1(\pi) = -\lambda^{2-\alpha}\sin (2\pi\alpha)/(\lambda^2+1)$,
yields the condition $\sin (2\pi\alpha) = 0$. Thus the determinant 
vanishes for $\alpha = i/2$,
where $i=1,2,3,\dots$, irrespective of the stretch $\lambda$ near C.
Among these possible solutions, the dominant
value of $\alpha$ for which the pressure is not singular at the cusp tip
is $\alpha = 5/2$, which means that the cusp opens with the universal
exponent $u \propto y^{3/2}$, as is confirmed over four decades
in Fig.~\ref{fig:crease} (b). Accordingly, $\alpha=5/2$ implies that 
the normal traction near the edge of the contact scales like
$t_n \propto |y|^{1/2}$, as confirmed numerically (inset). Thus,
both deformation and traction scale in the same way as a Hertz
contact \cite{J_book85}. 
We also note that the above calculation provides a rationale for the
scaling in the far field of the furrow: away from its rounded 
tip, the furrow's geometry is again slender and can described by
the same analysis.

Finally, we analyze the length $L$ of the fold. Since the energy density
of the fold solution (\ref{SP}) is constant, the contribution from the
fold is $E_f = A_0 L^2$, since $L$ sets the size of the area over which
deformation is significant. We can assume that the energy $E_0(d)$ of the
rest of the strain field is a smooth function of the deformation $d$.
Hence, if the creation of the fold takes place in a reversible fashion, 
we have $A_0 L^2 + E_0(d) = const$. Expanding $E_0$ linearly about $d=d_0$,
where $L = 0$, we obtain (\ref{L}). Apart from the experiment of
Fig.~\ref{fig:states}, this scaling law is confirmed with great precision
by the numerical simulation of a neo-Hookean material shown in
Fig.~\ref{fig:crease}~\cite{sup}.

In conclusion, our liquid-inclusion experiments allowed us to investigate
quantitatively localized furrows and creases which form on the surface
of an elastic medium under compression, and to document the hysteretic
transitions between them. We are able to describe the self-similar  shapes
of these furrows quantitatively, in agreement with both experiment and
neo-Hookean non-linear elasticity. Based on elasticity theory, we
are able to explain the $x\propto y^{3/2}$ scaling of the width of
both furrow and crease. 
These scaling laws reveal that the elastic singularity is a ``true" geometric cusp, 
and belongs to the same universality class as caustics in optics and free surface flows.

\acknowledgments{\paragraph{Acknowledgments.---} We are grateful to J. Dervaux and
L. van Wijngaarden for discussions. SK, AP and JHS acknowledge financial
support from ERC (the European Research Council) Consolidator Grant No. 616918.
J.E.'s work was supported by Leverhulme Trust Research Project Grant
RPG-2012-568.}


\end{document}